\documentclass[twocolumn, pra, showpacs,superscriptaddress]{revtex4}%
\usepackage{amsfonts}
\usepackage{amsmath}
\usepackage{amssymb}
\usepackage{graphicx}
\usepackage{subfigure}
\usepackage{epstopdf}
\usepackage{multirow}%
\providecommand{\U}[1]{\protect\rule{.1in}{.1in}}

\begin{document}
\title{Dynamics of a coupled spin vortex pair in dipolar spinor Bose-Einstein condensates}
\author{Tiantian Li}
\affiliation{Institute of Theoretical Physics, Shanxi University, Taiyuan 030006, China}
\author{Su Yi}
\email{syi@itp.ac.cn}
\affiliation{State Key Laboratory of Theoretical Physics, Institute of Theoretical Physics,
Chinese Academy of Sciences, P.O. Box 2735, Beijing 100190, China}
\author{Yunbo Zhang}
\email{ybzhang@sxu.edu.cn}
\affiliation{Department of Physics and Institute of Theoretical Physics, Shanxi University,
Taiyuan 030006, China}

\begin{abstract}
The collisional and magnetic field quench dynamics of a coupled spin-vortex
pair in dipolar spinor Bose-Einstein condensates in a double well potential
are numerically investigated in the mean field theory. Upon a sudden release
of the potential barrier the two layers of condensates collide with each other
in the trap center with the chirality of the vortex pair exchanged after each
collision, showing the typical signature of in-phase collision for the
parallel spin vortex phase, and out-of-phase collision for the antiparallel
phase. When quenching the transverse magnetic field, the vortex center in the
single-layered condensate starts to make a helical motion with oval-shaped
trajectories and the displacement of the center position is found to exhibit a
damped simple harmonic oscillation with an intrinsic frequency and damping
rate. The oscillation mode of the spin vortex pair may be tuned by the initial
magnetic field and the height of the Gaussian barrier, e.g. the gyrotropic
motions for parallel spin vortex pair are out of sync with each other in the
two layers, while those for the antiparallel pair exhibit a
double-helix-structure with the vortex centers moving opposite to each other
with the same amplitude.

\end{abstract}

\maketitle

\section{Introduction}

The multilayered structure with stacked vortices are intensively studied in
the past few years, as their simple structure combined with highly non-trivial
dynamical properties make them fascinating objects of research
\cite{nanomagnetism1,nanomagnetism,multilay1,multilay2,multilay3,multilay4,
multilay5,multilay6,multilay7,multilay8,multilay9,multilay10}. They also
provide the possibility to store information by means of their chirality and
polarity. As an ideal platform, a dipolar spinor Bose-Einstein condensate can
form a coreless vortex spontaneously in a pancake-shaped trap as a result of
the anisotropy of the magnetic dipolar-dipolar interaction (MDDI)
\cite{Kawaguchi06,Santos06,Yi06,Machida07,Yi08,Ho10,Hubert}.  
At intermediate temperature, a time-dependent Hartree-Fock-Bogoliubov study 
\cite{A1,A2} has shown that an anomalous vortex pair may be spontaneously 
generated in the anomalous fraction with the condensed atoms filled the vortex core, 
when phases corresponding to the singly charged vortex are imposed in the 
condensed and the anomalous components of the gas.
In particular, if an axial gaussian barrier is imposed upon the pancake-shaped
trap, the condensate will be divided into two layers with parallel or
antiparallel vortex in them for different barrier height \cite{Li}, which
resembles the trilayer films with a ferromagnetic-nonmagnetic-ferromagnetic
(F-N-F) structure.

In fact, the collisional dynamics of two topological defects in the condensate
has drawn great attention in recent years \cite{Kinoshita,Nguyen,Yang,
half1,half2,half3,skyrmions}. A quantum version of Newton's cradle has been observed in the
arrays of trapped one-dimensional (1D) bose gases by turning off the crossed
dipole trap quickly, which is in the proximity of an integrable system
\cite{Kinoshita}. Recently, an experiment on the collision of matter-wave
solitons consisted of a degenerate gas of $^{7}$Li atoms has been explored by
using a Gaussian laser beam to cut the condensate in half and then turning off
the barrier quickly \cite{Nguyen}. This would enable us to study collisions in
a more controlled manner and provides an ability to further explore the
transition between integrable and non-integrable systems. Another experimental
study \cite{half1,half2} on the collisional dynamics of two half-quantum vortices in
a highly oblate $^{23}$Na BEC is performed by means of a vortex-dipole
generation technique based on a moving optical obstacle, which demonstrates
the short range interaction between half-quantum vortices with different core
magnetizations. To understand the collisional dynamics of a spin vortex pair
\cite{Li} formed in the pancake-shaped dipolar spinor condensate needs a full
three-dimensional numerical simulation, which may provide further insight into
the face-to-face collisional dynamics of the spin vortex through the
controlled formation of parallel or antiparallel vortex pair in the two layers
of condensate.

A more practical motivation of this study lies in that magnetic vortex
oscillators have been investigated for use as tunable microwave generators for
future wireless communication \cite{Pribiag07,Dussaux10,Ruotolo09}. Hence a
better understanding of the fundamental physical properties of multilayered
vortex structure is urgent in order to improve their functionality. Recently,
investigations of dynamically coupled magnetic vortices have been reported in
pairs \cite{Sugimoto11,Jain12,Vogel11,Jung11,Shibata03,Sukhostavets}, trios
\cite{Wang}, chains \cite{Sukhostavets,Han}, and arrays
\cite{Sukhostavets,Vogel10,Barman10,Hanze14,Shibata14} of microscale
ferromagnetic disks and squares
\cite{Yu15,Choe,Guslienko2005,Waeyenberge,Yamada,Jun,Sluka,Phatak}. The
spontaneous parallel and antiparallel spin vortex pairs provide an ideal
platform for simulating multilayer magnetic vortices and a transverse magnetic
field may be applied to displace the vortex centers \cite{Li}, to achieve the
similar spin structure in nanopillar devices \cite{Sluka}. Moreover, the
coupled motion of vortices in a trilayer structure has been addressed by
time-resolved imaging technique with a scanning transmission x-ray microscope
\cite{multilay2,Jun}, which stimulates us to reveal the vortex dynamics in
coupled layers of the condensate after quenching the transverse field to zero.
\begin{figure*}[ptb]
\includegraphics[width=0.95\textwidth]{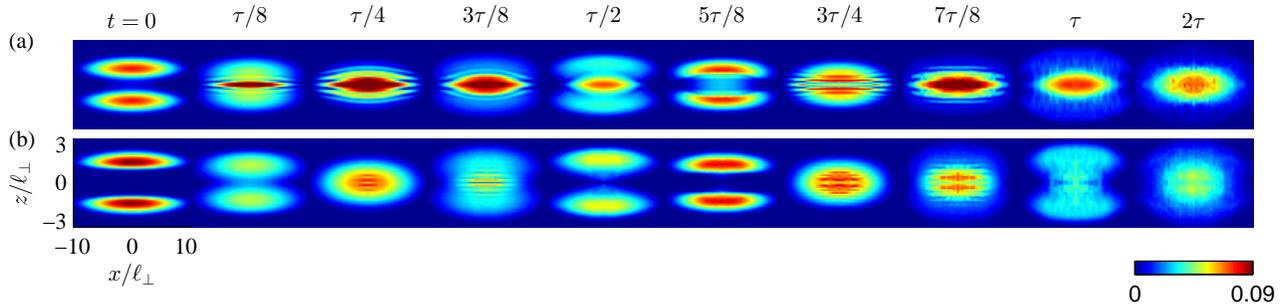}\centering
\caption{(Color online) Collision dynamics of a coupled spin vortex pair in
(a) PSVP ($A = 100\hbar\omega_{\perp}$) and (b) ASVP ($A = 300\hbar
\omega_{\perp}$) phases of the condensate when the Gaussian barrier is turned
off suddenly. Each image denotes the cloud density in $x-z$ plane after
integrating along $y$ axis.}%
\label{fig1}%
\end{figure*}

In consideration of the wide applications of magnetic vortex, in this paper we
focus on the collisional and quench dynamics of the initial spin vortex ground
states in Ref. \cite{Li} by suddenly turning off either the potential barrier
of the double well or the transverse magnetic field. The paper is organized as
follows. In Sec.~\ref{form}, we introduce our model and briefly review the
ground state properties of this system, which is necessary to understand the
following discussions. The numerical results on the collisional dynamics are
presented in Sec.~\ref{rmbarr}. In Sec.~\ref{rmmag} we study the magnetic
field quench dynamics of the spin vortex for different phases. Finally, we
summarize our findings in Sec.\ref{concl}.

\section{Formulation}

\label{form}

The dynamics of the spin vortex pair formed in a dipolar spinor condensate is
governed by the following set of Gross-Pitaevskii equations in the mean-field
treatment
\begin{equation}
i\hbar\frac{\partial\psi_{\alpha}}{\partial t}=\left(  T+U+c_{0}n\right)
\psi_{\alpha}+g_{F}\mu_{B}{\mathbf{B}}_{\mathrm{eff}}\cdot{\mathbf{F}}%
_{\alpha\beta}\psi_{\beta}, \label{gpe}%
\end{equation}
where $\psi_{\alpha}({\mathbf{r}})$ denotes the condensate wave function for
the spin component $\alpha=0,\pm1$ respectively, $n({\mathbf{r}})=\sum
_{\alpha}|\psi_{\alpha}|^{2}$ is the total density, $T=-\hbar^{2}\nabla
^{2}/(2M)$ is the kinetic energy with $M$ the atomic mass, and $U\left(
\mathbf{r}\right)  $ is the trapping potential. The coefficient $c_{0}=4
\pi\hbar^{2} \left(  a_{0}+2a_{2}\right)  /(3M)$ describes the
spin-independent collisional interaction in the condensate with $a_{0,2}$
being the $s$-wave scattering length for two spin-1 atoms in the total spin
channels $0$ and $2$, $g_{F}$ is the Land\'{e} $g$-factor of the atom,
$\mu_{B}$ is the Bohr magneton, and ${\mathbf{F}}$ is the angular momentum
operator. All spin related interactions are collected into the last term in a
way of an effective magnetic field
\begin{align}
{\mathbf{B}}_{\mathrm{eff}}({\mathbf{r}})  &  ={\mathbf{B}}_{\mathrm{ext}%
}+\frac{c_{2}}{g_{F}\mu_{B}}{\mathbf{S}}({\mathbf{r}})\nonumber\\
&  +\frac{c_{d}}{g_{F}\mu_{B}}\int\frac{d{\mathbf{r}}^{\prime}}{|{\mathbf{R}%
}|^{3}}\left[  {\mathbf{S}}({\mathbf{r}}^{\prime})-\frac{3\left[  {\mathbf{S}%
}({\mathbf{r}}^{\prime})\cdot{\mathbf{R}}\right]  {\mathbf{R}}}{|{\mathbf{R}%
}|^{2}}\right]  \label{beff}%
\end{align}
with ${\mathbf{R}}={\mathbf{r}}-{\mathbf{r}}^{\prime}$ and ${\mathbf{B}%
}_{\mathrm{ext}}$ the external magnetic field. Here, ${\mathbf{S}}%
({\mathbf{r}})=\sum_{\alpha\beta}\psi_{\alpha}^{*}{\mathbf{F}}_{\alpha\beta
}\psi_{\beta}$ is the spin density, and $c_{2}$ and $c_{d}$ characterize the
spin exchange and dipole-dipole interaction, respectively.

The dipolar interaction in the condensate induces interesting spin vortex
structure in the ground state for a pancake-shaped trap potential $U\left(
\mathbf{r}\right)  $. To create a spin vortex pair, a Gaussian barrier of
height $A$ and width $\sigma_{0}$ is imposed upon a highly anisotropic
harmonic potential with aspect ratio $\lambda$, leading to a combined trapping
potential $U\left(  \mathbf{r}\right)  =\frac{1}{2}M\omega_{\perp}^{2}\left(
x^{2}+y^{2}+\lambda^{2}z^{2}\right)  +Ae^{-z^{2}/(2\sigma_{0}^{2})} $ with
$\omega_{\perp}$ the radial trap frequency. In experiment, one may choose a
Gaussian laser beam to cut the condensate into two halves \cite{Nguyen} and
the system can be considered as quasi-two-dimensional clouds of atoms, nearly
free to move in the planes. The vortex pair could be with either parallel or
anti-parallel spin on the two sides of the barrier \cite{Li} and the spin
structures of the ground state denoted as the PSVP and ASVP are determined by
the height of the Gaussian barrier $A$ for a fixed barrier width $\sigma_{0}$
in the absence of an external magnetic field. The competition between the
tunneling splitting of the two wells and the interlayer MDDI energy induces a
phase transition between PSVP and ASVP. An increasing transverse magnetic
field is found to remove the spin vortices sequentially before the condensate
is fully polarized.

By preparing the system in the two distinct phases it is interesting to study
the dynamics of coupled spin vortex pair by turning off either the barrier or
the magnetic field, both of which are described by solving numerically the
time dependent GP equation Eq.~(\ref{gpe}) with the help of the real time
propagation method. To study the dynamics of the coupled vortex pair, the initial states for the quench 
dynamics of the condensate are prepared in the spin vortex state based on our 
previous work on the ground state of the system \cite{Li}.
For $^{87}$Rb, we take $c_{2}=0.01c_{0}$ and deliberately increase the DDI parameter to
$c_{d}=\left\vert c_{2}\right\vert $ to accelerate the convergence of the
numerical calculation as in Ref. \cite{Li}. In addition, we take the same parameters as in
previous study, i.e. total atom numbers $N=5\times10^{5}$, $\omega_{\perp
}=(2\pi)100\,$Hz, $\sigma_{0}=0.65\ell_{\perp}$, and $\lambda=6$ in our
numerical calculation, and the dimensionless units are $\hbar\omega_{\perp}$
for energy, $\ell_{\perp}=\sqrt{\hbar/(M\omega_{\perp})}$ for length,
$\omega_{\perp}^{-1}$ for time, and $\ell_{\perp}^{-3/2}$ for the wave
function, respectively. Enhancement of the DDI by increasing the number of 
particles only slightly changes the critical barrier height $A^{*}$ between the ASVP and PSVP phases. 
For example, for $N=5 \times 10^5$ adopted in this work, $A^{*}=246 \hbar \omega$, while for 
$N=2 \times 10^5$ we have $A^{*}=158 \hbar \omega$. This, however, does not change the dynamics 
qualitatively. In fact, it was verified that the qualitative results can be reproduced by solely 
increasing $N$ to $10^7$ with $c_d$ being that of the $^{87}$Rb atom. 

\section{Collision Dynamics}

\label{rmbarr}

In order to provide an experimentally discernible signature of the PSVP and
ASVP phases, we study the collision dynamics of the spin vortex pair by
turning off the barrier quickly in a very short time once the pair is formed
in the upper and lower planes, i.e. the system is prepared deliberately at the
beginning in the ground state of the double well with $A=100\hbar\omega
_{\perp}$ and $300\hbar\omega_{\perp}$, respectively. The atoms suddenly find
themselves at the classical turning points of the harmonic trap in the $z$
direction and begin to accelerate towards the centre. We carried out a full
three dimensional calculation of the collision dynamics and the integrated
density profiles are presented in Fig. \ref{fig1} for two oscillation cycles.
As a result, the Newton's cradle-like dynamics \cite{Kinoshita} is observed in
our system, featured with distinct interference fringes and chirality exchange
in the spin vortex pair.

\begin{figure}[ptb]
\includegraphics[width=0.95\columnwidth]{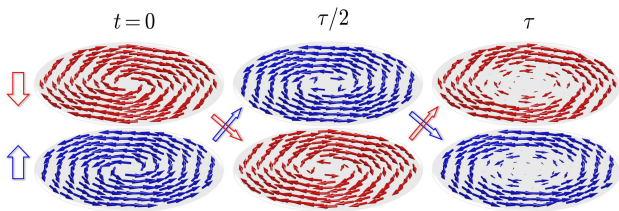} \centering
\caption{(Color online) Spin vortex pair in the three collisional moments for
the ASVP phase. Spin structures in the initial potential minima are shown in
the upper and lower layers with the chiralities exchanged after each collision
- the chiralities are inverted at $t=\tau/2$ and restored at $t=\tau$.}%
\label{fig2}%
\end{figure}

The two sheets of cloud collide with each other in the center of the trap
twice each full cycle, for instance at $t=\tau/4$ and $3\tau/4$ with
$\tau=2\pi/\omega_{z} =1.05 \omega_{\perp}^{-1} $, as illustrated in Fig.
\ref{fig1}. When they first meet, the atoms in the PSVP phase produces global
interference patterns in the whole condensate, while those in the ASVP phase
are stirred only in the central area. At this moment, the collisional
interaction among the three components is the strongest so that the exchange
of the particle number reaches to most frequently. We find that the density of
spin-0 component $n_{0}$, defined as the full space integration of $\left\vert
\psi_{0}\right\vert ^{2}$, decreases during the first collision time for the
PSVP phase, indicating that this phase favors a particle exchange direction
$2\left\vert 0\right\rangle \rightarrow\left\vert 1\right\rangle +\left\vert
-1\right\rangle $, while it increases for the ASVP phase and an opposite
particle exchange channel is preferred. It is worthwhile to mention that the
magnetization of $z$ axis $M_{z}=\int d\mathbf{r} ( \left\vert \psi
_{1}\right\vert ^{2}-\left\vert \psi_{-1} \right\vert ^{2} )$ remains zero in
the whole evolution process so that the particle number density of $\pm$
components always satisfies $n_{1}=n_{-1}$.

And then, the two layers of condensate pass straightly through each other
before reemerging on the other side. When the condensates separate into two
parts again at $t=\tau/2$, we observed an interesting phenomenon that the
chiralities of the spin vortex pair in the upper and lower layers, i.e. the
circulations of the in-plane magnetization components colored by red and blue
in Fig. {\ref{fig2}}, exchange with each other in the ASVP phase and are
restored at $t=\tau$ except that the spin structures are a little distorted
near the vortex center, while those in the PSVP phase are the same as they
should be. The phase differences between the lower and upper parts $\phi_{0}$
(which is $0$ in the PSVP and $\pi$ in the ASVP) leads to drastically
different behavior in the wave functions: at the half period $t=\tau/2$ a
density peak appears at the centre-of-mass of the condensate for the PSVP
indicating an in-phase collision, which is not observed for the ASVP case
implying an out-of-phase collision. These phenomena are in good agreement with
the experimental observation of the collision of matter wave solitons
\cite{Nguyen}. Different with the 1D collision, the face-to-face oscillation
here in two dimensional will be destructed in a few cycles, which can be
understood as a result of the density-dependent inelastic collision.

\begin{figure}[ptb]
\includegraphics[width=0.95\columnwidth]{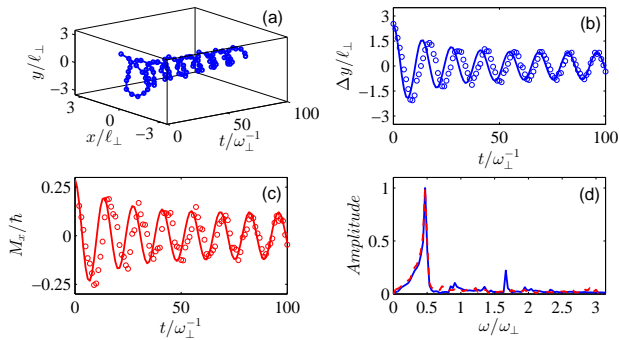} \centering
\caption{(Color online) Time-resolved trajectories of the vortex centers in
the $z=0$ plane (a) and the displacement of the vortex center in $y-$axis (b)
when the transverse magnetic field ($B_{x}=25\mu G$) is removed suddenly after
initially preparing the system in the ground state of a harmonic trap. The
magnetization $M_{x}$ in (c) shows similar oscillation behavior with the same
characteristic frequency as can be seen from the spectral analysis in (d) for
the two signals. The solid lines in (b) and (c) are fitted with a damped
simple harmonic oscillation. }%
\label{fig3}%
\end{figure}

\section{Quench Dynamics of the Transverse Magnetic Field}

\label{rmmag}

The dynamics of magnetic vortices in nanodisks or multilayer structures
manipulated by the magnetic field has attracted more and more attention
\cite{Jun,Choe,Guslienko2005,Yu15,Sluka}. The rotation of a vortex as a whole
around its equilibrium position is found in a gyrotropic mode with a frequency
far below the ferromagnetic resonance of the corresponding material
\cite{Jun}. Here we show similar oscillation mode of the spin vortex pair
tuned by the magnetic field in the dipolar spinor condensate system. We focus
on the motion of the vortex center in three typical structures, i.e. the
single layer, PSVP and ASVP, when the transverse field is suddenly quenched to
zero. In the following we show the time-resolved trajectories of the vortex
centers in the $z=z_{min}$ plane with $z_{min}$ being the positions of the
potential minima along the $z$ axis. In the three cases discussed here, we
have $z_{min}=0$ for single layer condensate in a harmonic trap, and
$z_{min}=\pm\sigma_{0} \sqrt{2 \ln\left(  A/\lambda^{2}\sigma_{0}^{2}\right)
}$ for the double well structure.

\textit{Dynamics of a spin vortex in a single-layered condensate}: We first
consider the motion of the vortex center in a single-layered, pancake-shaped
condensate starting from an initial state with a spin vortex for $B_{x}=25\mu
G$ which may be regarded as a reference for more complex structure. As
demonstrated in Fig. \ref{fig3}(a), when the field is removed quickly, the
vortex center in the $z=0$ plane starts to make a helical motion with
oval-shaped trajectory in $xy$ plane. The projection of the center position on
the $y-$axis, that is the displacement of the vortex center, is found to
exhibit a well-defined periodic oscillation for quite a long time, as shown in
Fig. \ref{fig3}(b). This gyrotropic motion of the vortex center has been
fitted with a damped simple harmonic oscillation motion as $\Delta
y(t)=\left(  y_{0} \exp(-\beta t)+y_{1}\right)  \cos(\omega_{0} t)$, where
$y_{0}=1.7 \ell_{\perp}$, $y_{1}=0.8 \ell_{\perp}$, $\beta= 5.84 \times
10^{-2}\omega_{\perp}$, and $\omega_{0}=0.46 \omega_{\perp}$. This can be
explained by taking into account the fact that the magnetic vortex is driven
to move by a precessional torque generated by turning off the transverse
magnetic field suddenly. We also examine the time dependence of the
magnetization along $x$ axis $M_{x}=\int d{\mathbf{r}}S_{x}({\mathbf{r}})$ in
Fig. \ref{fig3}(c), the oscillation there has been fitted with a formula
$M_{x}(t)=\left(  M_{0} \exp(-\beta t)+M_{1}\right)  \cos(\omega_{0} t)$ with
$M_{0}=0.17 \hbar$ and $M_{1}=0.12 \hbar$. We find that the characteristic
damping parameter $\beta$ and the oscillation frequency $\omega_{0}$ match
exactly with those in the evolution of displacement of the vortex centers
$\Delta y(t)$. This can be clearly seen in the Fourier spectral analysis of
the two oscillations in Fig. \ref{fig3}(d). During the evolution, however, the
magnetizations along $y$ and $z$ axis are found to be constantly zero.

\begin{figure}[ptb]
\includegraphics[width=0.95\columnwidth]{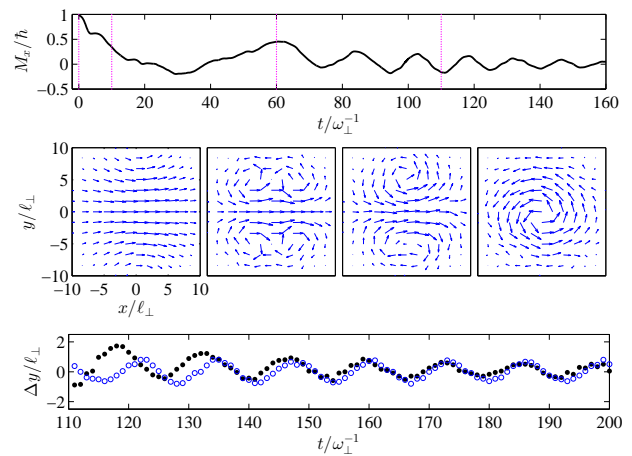}
\centering
\caption{(Color online) Quench dynamics of the spin structure in a
single-layered condensate staring from a polarized ground state with a
transverse magnetic fields $B_{x} = 70\mu G$. The top panel shows the time
dependence of the magnetization along $x$ axis. The middle panels show
formation and stabilization of the spin vortex in the $z=0$ plane at four
evolution times $t= 0, 10, 60$ and $110 \omega^{-1}_{\perp}$ (magenta dotted
vertical lines in the upper panel), respectively. The filled circles in the
lower panel shows the displacement of the vortex center in $y$ axis after the
stabilization of the spin vortex, and the open circles show the corresponding
long time oscillation of the vortex center for $B_{x}=25\mu G$ in Fig.
{\ref{fig3}b}.}%
\label{fig4}%
\end{figure}

Staring from a polarized state gives another scenario of the quench dynamics
of the spin structure in a single-layered condensate, including the vortex
formation and its helical motion. Fig. \ref{fig4} shows the details: four
vortices are formed after the quench of the magnetic field, and they merged
sequentially into two, and then to a single vortex. Meanwhile the
magnetization is found to undergo a rather slow damping oscillation. Once the
vortex is stabilized, its center moves in just the same way as the helical
motion with oval-shaped trajectory, the frequency of which tends to match the
small transverse field case for $B_{x}=25\mu G$ as shown in the bottom panel
in Fig. \ref{fig4}. We therefore conclude that the oscillation frequency of
the vortex center in the single-layered condensate is intrinsic and irrelevant
to the quenching initial state.

\begin{figure}[t]
\includegraphics[width=0.95\columnwidth]{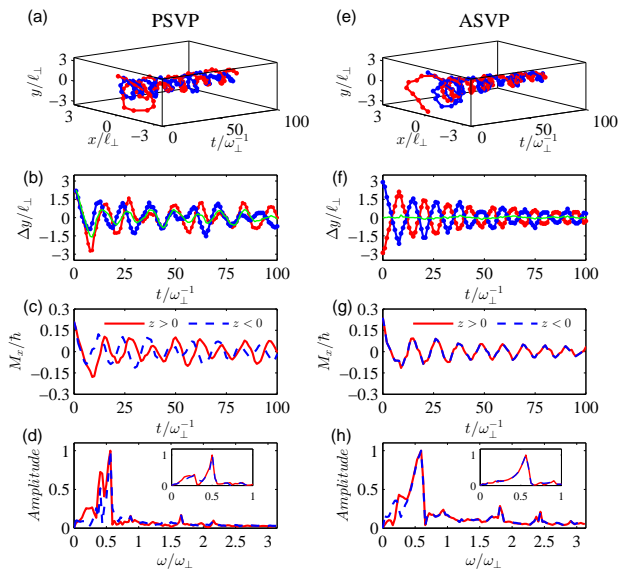} \centering
\caption{(Color online) Time-resolved trajectories of the vortex centers in
the $z=z_{min}$ planes (the first row) and the displacement of the vortex
center on $y$-axis (the second row) when the transverse magnetic field
($B_{x}=25 \mu G$) is removed suddenly after initially preparing the system in
the PSVP (left) and ASVP (right) phases, respectively. Panels in the third row
show the time evolution of the corresponding magnetizations $M_{x}$ in the
upper ($z>0$) layer and lower layer ($z<0$), both of which oscillate in the
same characteristic frequency as the motion of the vortex centers. The
spectral analysis for the two signals are plotted in the bottom panels while
the insets show the situation for a wider separation of the two layers
$\sigma_{0}=1.2 \ell_{\perp}$. The red (blue) lines are for the upper (lower)
layer and the green lines in the second row plot the average position of the
two vortex centers.}%
\label{fig5}%
\end{figure}

\textit{Dynamics of a spin vortex pair in the PSVP and ASVP phases}: For the
PSVP or ASVP ground state, the number of spin vortices decreases from 2 to 0
with a growing transverse field in the three stages of the magnetization
process, i.e. double-vortex, single-vortex, and polarized states. We first
examine a small transverse field with $B_{x}=25\mu G$, in which case the spin
vortex pair is formed in two sheets of condensate with the same chiralities
and the centers displaced along the same direction for the PSVP state, and
with opposite chiralities and the two centers separated away along opposite
directions for the ASVP state. Quenching this transverse field abruptly to
zero, we find that the dynamics of the vortex centers is much different from
that of a single-layered condensate due to the existence of a significant
interplay between the interwell MDDI and tunneling splitting of the
double-well potential. For the PSVP phase, the vortex centers of the two
layers make a gyrotropic motion but out of sync with each other, as shown in
Fig. \ref{fig5}(a) and their projection on $y$ axis in Fig. \ref{fig5}(b).
Clearly the average position of the two vortex centers is not zero due to the
frequency difference of the oscillations in the upper and lower layers (the
green line in Fig. \ref{fig5}(b)). We also examine the time dependence of the
magnetization $M_{x}$ in Fig. \ref{fig5}(c), where the magnetization in the
two parts oscillate in the same way as the displacement of the vortex centers,
just as what happens in the single-layered condensate. In particular, we make
a Fourier transform of the magnetization dynamics over a period of
$100\omega_{\perp}^{-1}$ and find that for each layer a beat phenomenon
appears in the dynamics with two slightly different characteristic frequency
peaks as shown in Fig. \ref{fig5}(d). Further numerical simulations show that
the weight of two frequencies in the dynamics changes along with the width of
the Gaussian barrier. As the distance between the layers increases, the
principle peak moves to higher frequency which represents the gyrostropic
vortex motion, while the other peak corresponds to its higher harmonics. This
can be understand as following: Note that both the interwell dipolar
interaction (proportional to $1/\sigma_{0}^{3}$) and tunneling splitting
(roughly proportional to $e^{-\widetilde A\sigma_{0}}$) decrease when the two
layers are separated wider at a fixed height of the interwell barrier
$\widetilde A$ which is measured from the minima of the double-well potential.
In the PSVP phase, the dipolar interaction inside each layer thus plays a
major role so that the dynamics is dominated by a relatively strong nonlinear
interaction as in the single layer case with the principle characteristic
frequency more prominently located at $0.50 \omega_{\perp}$ for $\sigma
_{0}=1.2 \ell_{\perp}$ as can be seen in the inset of Figure \ref{fig5}(d).

As forming an ASVP state needs a higher barrier $A$, the tunneling interaction
between the two wells will be weakened so that the interwell dipolar
interaction plays a key role in the process of the evolution. The trajectories
of the vortex centers in ASVP phase are shown in Fig. \ref{fig5}(e) and their
projections in Fig. \ref{fig5}(f). We find that starting from two opposite
positions in the $xy$ plane, the two vortices with opposite chiralities
exhibit a double-helix-structured gyrotropic motion and exhibit damped
oscillation behaviors. It is worthwhile to notice that the motion of the
vortex centers in the two layers are opposite to each other, but with the same
amplitude, leading to a nearly zero average displacement of the vortex centers
(the green line in Fig. \ref{fig5} (f)). The magnetization $M_{x}$ in each
layer shows perfect match during the damped evolution (Fig. \ref{fig5}(g)).
Obviously the beat phenomenon exists neither in the motion of vortex cores nor
the evolution of the magnetization $M_{x}$ (Fig. \ref{fig5}(f) and (g)) just
as in the single layer case. The Fourier transform of the oscillation is a
single peaked spectrum with the characteristic frequency located at
$0.56\omega_{\perp}$ for $\sigma_{0}=1.2\ell_{\perp}$ as shown in the inset of
Fig. \ref{fig5}(h). The damping rate $\beta$ of the oscillation can be
adjusted with a varying barrier separation $\sigma_{0}$, for instance it is
three times faster for a closer space $\sigma_{0}=0.5\ell_{\perp}$ for which
we have $\beta=0.03\omega_{\perp}$ than a wider separation $\sigma_{0}%
=1.2\ell_{\perp}$ for which we have $\beta=0.01\omega_{\perp}$ with the
fitting frequency $\omega_{0}=0.52 \omega_{\perp}$.

\begin{figure}[t]
\includegraphics[width=0.95\columnwidth]{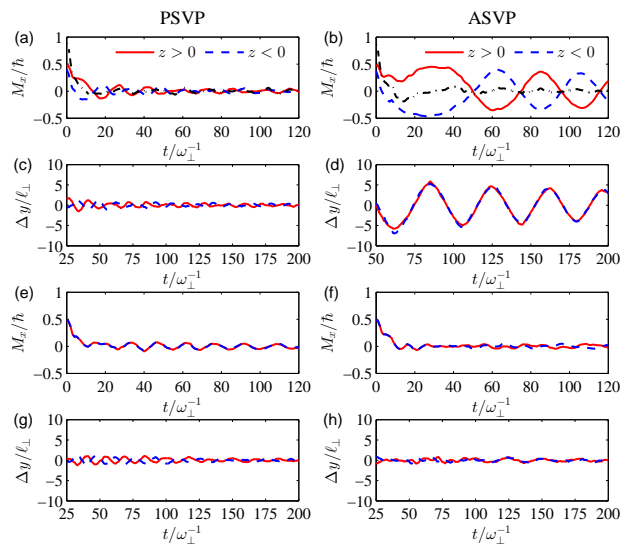} \centering
\caption{(Color online) Dynamics of a single vortex and a polarized state in
the PSVP and ASVP. The time dependences of the magnetization along $x$ axis
are shown in (a), (b), (e) and (f) and the displacement of the vortex center
on $y$ axis in (c), (d), (g) and (h) for the PSVP case (left) and ASVP case
(right), respectively, when the magnetic field is turned off suddenly. The
corresponding initial transverse magnetic fields are $B_{x}=60 \mu G$ for (a)
and (c), $50\mu G$ for (b) and (d), and $70\mu G$ for (e-h).}%
\label{fig6}%
\end{figure}

\textit{Dynamics of a single vortex in the PSVP and ASVP}: Next, we choose an
intermediate magnetic field such that the initial state has a single vortex
in, say, the lower layer, while the upper layer is polarized with $M_{x} (z<0)
< M_{x}(z>0)$. Depending on the magnetic phases of the condensate, the system
can support different combinations of vortex states. We compare the results
for PSVP and ASVP in Fig. \ref{fig6} which shows the time dependence of the
magnetization $M_{x}$ of the condensate for $A=100$ and $300\hbar\omega
_{\perp}$, $\sigma_{0} =0.65\ell_{\perp}$ after quenching the initial magnetic
field $B_{x}=50 \mu G$ for PSVP, and $B_{x}=60 \mu G$ for ASVP, to zero. As we
can see, the evolution here is similar to the single layer case and can be
divided into two stages: (i) vortex formation stage - compared to the
stabilization of the PSVP spin vortex pair with the same chirality, it takes a
little longer time for the two layers in the ASVP phase to form a spin vortex
pair with opposite chiralities, and (ii) vortex oscillation stage - the
centers of the stabilized vortices start to make the typical helical motion
after $t=25 \omega^{-1}_{\perp}$ for PSVP and $t=50 \omega^{-1}_{\perp}$ for
ASVP. We note that both the magnetization oscillation and the displacement of
the vortex center are rather small for PSVP phase, while those in the ASVP
phase show quite a larger amplitude and lower frequency. Also, the two vortex
centers in the ASVP phase follow the same trajectory as in Fig. \ref{fig6}(d),
while the magnetizations in each layer are opposite to each other leading to a
nearly vanishing total magnetization as indicated by the black dash-dotted
line in Fig. \ref{fig6}(b).

\textit{Dynamics of a polarized state in the PSVP and ASVP:} At last, we
consider a polarized initial state in which all spins are polarized along $x$
direction at a large magnetic field $B_{x}=70 \mu G$. The fate of this state
in a long time is a spin vortex pair with opposite chiralities, irrelevant
with the initial PSVP or ASVP phases. In this regime the magnetization in the
two layers show uniform behavior in both PSVP and ASVP phases, and in Fig.
\ref{fig6}(e-h) we still observe that the magnetization and the vortex centers
of each layer oscillate with time for PSVP, in the ASVP phase, however, the
oscillation is greatly suppressed, i.e. the vortex center hardly move with
their centers pinned at the origin of the $xy$ plane once the vortices are
stabilized. These features may be exploited to identify the individual
configurations of PSVP and ASVP experimentally.

\section{CONCLUSION}

\label{concl}

Before concluding, we would like to point out the connections of this study to 
the existing experimental works in some related fields. The appearance and dynamics 
of vortex and half-vortex(HV) in polariton condensate \cite{half3} and spin-1 
anti-ferromagnetic condensate \cite{half1} are connected with spontaneous 
symmetry breaking and phase transitions. The cores of the primary half-vortex in 
the polariton condensate are seen moving along orbital-like trajectories around the 
density maximum of the counterpolarized Gaussian state, keeping itself orbiting 
during a few tenths of picoseconds, while the trajectories of the twin singularities 
upon full-vortex injection undergo a spiraling at large polariton densities. 
Such dynamical configuration resembles the gyrotropic motions of the vortex center 
by manipulating the transverse field in double-layered dipolar condensate (see Fig. (\ref{fig5}))and 
ferromagnetic nanodisks \cite{Guslienko2005,Jun}. By means of in situ 
magnetization-sensitive imaging, pairs of half-quantum vortices with opposite core 
magnetization are generated when singly charged quantum vortices are injected into 
the easy-plane polar phase of an antiferromagnetic spinor Bose-Einstein condensate \cite{half1}. 
The separation distance of the pair and the magnitude of the core magnetization gradually 
saturate that is consistent with the short-range repulsive interactions between half-quantum 
vortices with opposite core magnetization. In our system, both the displacement of the vortex 
center and the magnetization along $x$ axis exhibit well-defined periodic oscillation, 
which is attributed to the long-range dipole-dipole interaction in the condensate.

In conclusion, we have investigated the collisional and magnetic field quench
dynamics of the coupled spin vortex pair formed in a weakly interacting
dipolar spinor BEC in a double well potential. We show that, in the absence of
a magnetic field, the two sheets of the cloud collide with each other with
their chiralities exchanged after each collision, when the potential barrier
is turned off suddenly. Especially, the evolutions of the vortex pair in PSVP
and ASVP phases can be easily distinguished from the density distribution
images at half of the oscillation period $\tau/2$ where the in-phase and
out-of-phase collisions differ markedly depending on the relative phase
between the upper and lower layers. Motivated by the experimental studies of
the dynamics of magnetic vortices in nanodisks or multilayer structures, we
subsequently study the quench dynamics of the vortex in the ground states of a
single- and double-layered condensate when the initial transverse field is
removed quickly. The motion of the vortex center is found in a gyrotropic mode
on account of the spin torque generated by quenching the transverse field. The
displacement of the vortex center and the corresponding transverse
magnetization are fitted with a damped simple harmonic oscillation motion with
an intrinsic frequency and damping rate. These interesting dynamics provide an
experimentally discernible signature of the PSVP and ASVP phases in the reach
of current techniques of cold atom experiments. In future works we intend to
develop the non-zero temperature theory to understand the effect of spin fluctuations 
and thermal magnons on the dynamics.

\begin{acknowledgments}
This work was supported by the NSFC (Grant Nos. 11234008 and 11474189) and by
the National 973 Program (Grant No. 2011CB921601), the Open Project Program of
State Key Laboratory of Theoretical Physics, Institute of Theoretical Physics,
Chinese Academy of Sciences, China (No.Y5KF231CJ1). SY acknowledge the support
from NSFC (Grants Nos. 11434011 and 11421063).
\end{acknowledgments}

\end{document}